\documentclass[prb,twocolumn,aps,showpacs,fixfloats]{revtex4}
\usepackage{graphicx}
\usepackage{subfigure}
\usepackage{float}
\usepackage{bm}
\usepackage{amsmath,amssymb}
\usepackage{latexsym}
\usepackage{color}
\usepackage{enumerate}
\usepackage{pdfpages}
\usepackage{tikz}
\usepackage{hyperref}

\begin{document}
\newcommand{\s}{\scriptscriptstyle}
\newcommand{\uu}{\uparrow \uparrow}
\newcommand{\ud}{\uparrow \downarrow}
\newcommand{\du}{\downarrow \uparrow}
\newcommand{\dd}{\downarrow \downarrow}
\newcommand{\ket}[1] { \left|{#1}\right> }
\newcommand{\bra}[1] { \left<{#1}\right| }
\newcommand{\bracket}[2] {\left< \left. {#1} \right| {#2} \right>}
\newcommand{\vc}[1] {\ensuremath {\bm {#1}}}
\newcommand{\tr}{\text{Tr}}
\newcommand{\Trans}{\ensuremath \Upsilon}
\newcommand{\Refl}{\ensuremath \mathcal{R}}

\title{Spinful Aubry-Andr{\'e} model in a magnetic field: Delocalization facilitated  by a weak spin-orbit coupling}

\author{Rajesh K. Malla   and M. E. Raikh}

\affiliation{ Department of Physics and
Astronomy, University of Utah, Salt Lake City, UT 84112}

\begin{abstract}
We have incorporated spin-orbit coupling into the Aubry-Andr{\'e} model
of tight-binding electron motion in the presence of periodic potential
with a period incommensurate with lattice constant. This model is known
to exhibit an insulator-metal transition upon increasing the hopping amplitude.
%potential.
Without external magnetic field, spin-orbit coupling
%this coupling
leads to a simple renormalization
of the hopping amplitude. However, when the degeneracy of the on-site energies is
lifted by an external magnetic field, the interplay of Zeeman splitting and
spin-orbit coupling has a strong effect on the localization length.
%delocalizing effect.
%effect on the localization transition in this model.
We studied this interplay numerically
%in the semiclassical limit,
by calculating the energy dependence of the Lyapunov exponent in the
insulating regime. Numerical results can be unambiguously 
%The results are 
interpreted
%the numerical results
in the language of the phase-space trajectories.
As a first step, we have explained the plateau in
the energy dependence of the localization
length in the original Aubry-Andr{\'e} model.
Our main finding is that a very weak spin-orbit coupling leads to
delocalization of states with energies smaller than the Zeeman shift.
The origin of the effect is the spin-orbit-induced opening of new transport channels.
We have also found that restructuring of the
phase-space trajectories, which takes place at certain energies {\em in the insulating regime},
causes a singularity in the energy dependence of the localization length.
%If the period of incommensurate potential contains more than $\sim 10$ sites,
%then the semiclassical description applies, and the localization length becomes a
%smooth function of energy. We have established the singularity in the energy dependence
%of the localization radius inside the insulator regime. This singularity reflects the
%restructuring of the phase-space trajectories, which takes place at certain critical
%spin-orbit coupling strength.
\end{abstract}

\pacs{73.50.-h, 75.47.-m}
\maketitle

\section{Introduction}

%\begin{figure}
%\label{ft1smallperiod}
%\includegraphics[scale=0.24]{t10smallperiod.pdf}
%\includegraphics[scale=0.24]{t1smallperiod.pdf}
%\caption{(Color online) The Lyapunov exponent as a function
%of energy is plotted for canonical $\beta=\frac{\sqrt{5}-1}{2}$
%in the insulating regime, $t=0.4V$,   for Zeeman splitting $\Delta=V$.
%In the upper panel, (a), spin-orbit coupling
%is absent. In the lower panel, (b), the parameter $t_1$ is chosen to be
%$t_1=0.025V$. It is seen that plots differ near $E=0$, where small
%$t_1$ causes the metallization. Blue and red curves in (b) correspond to two
%eigenvalues of the transmission matrix.}
%\end{figure}

A standard description of electron motion in a one-dimensional quasiperiodic potential
is based on the Aubry-Andr{\'e} (AA) model\cite{Original} with tight-binding Hamiltonian
\begin{equation}
\label{eqh0}
\hat{H}_0=-t\sum\limits_n \left(c_{n}^{\dagger}c_{n+1}+ c_{n+1}^{\dagger}c_{n}\right)+V\sum\limits_n\cos(2 \pi \beta n)c_{n}^{\dagger}c_{n},
\end{equation}
where $c_n^{\dagger}$ is the creation operator of electron on $n$-th site,  $t$ is the
hopping integral, $V$ is the amplitude of modulation of the on-site energies, and $\beta^{-1}$ is the modulation period. Non-triviality of the AA model originates from
the fact that, for irrational $\beta$, it exhibits a delocalization transition and yet contains no
randomness.
%disorder.

The key finding of Ref. \onlinecite{Original} is that the Hamiltonian Eq.~(\ref{eqh0})
possesses self-duality: upon transformation from coordinate to the momentum space
it retains its form after the interchange
$V\hspace{-0.7mm} \leftrightarrow \hspace{-0.7mm} 2t$. The consequence of this duality\cite{SokoloffReview} is
that, for $V>2t$, all eigenstates are exponentially localized with localization length scaling as $(V-2t)^{-1}$. From the perspective of physics, the importance of the AA model
is that it captures the peculiarities of motion of a two-dimensional electron in a perpendicular magnetic field and a periodic potential.\cite{Hofstadter,Azbel1979,Thouless1982}

Early studies of the AA model
\cite{Sokoloff1980,Suslov,Soukoulis,Thouless1983,Kohmoto,Kohmoto1}
were focused on the properties of localized eigenfunctions near the transition.
Lately, the interest to the AA model has been revived
\cite{DasSarmaEdges,InOpticalLattice,Flach2013,
KaiSun+DasSarma2013,Shlyapnikov,Interacting2015,
InOpticalLattice1,QuenchDynamics,AnomalousDiffusion,SO+bosons,
NigelCooper,ac-driven,Skipetrov,QuenchDynamics1,AAAdiabaticPumping}.
Nowadays, it is invoked to study different observable quantities
in the presence of the quasiperiodic background.
These studies were largely motivated by two groups of
experiments Refs. \onlinecite{Experiment2008,Experiment2015,Experiment2017} and Refs. \onlinecite{Experiment2009,Experiment2012,Experiment2013}.
%\onlinecite{Experiment2012}.
In  the experiment Ref. \onlinecite{Experiment2008} the expansion
of cold atoms loaded into an optical lattice was studied. One-dimensional
modulation was formed as a result of interference
of two laser beams. Localization transition, which takes place upon
increasing the modulation amplitude, was demonstrated through the
analysis of spatial and momentum distribution of atoms released
from the lattice. In Refs. \onlinecite{Experiment2015}, \onlinecite{Experiment2017}
the degree of localization of cold fermions in quasiperiodic optical lattice
was monitored
via the time evolution of the imbalance of population of
different sites following a quench of system parameters.

In experiments of the second group\cite{Experiment2009,Experiment2012,Experiment2013},
propagation of light along the axes of coupled waveguides
has been studied. The centers of waveguides formed a periodic array,
while their parameters were periodically modulated.
Localization transition has been detected
via the spreading of initially narrow wave
packet across  the lattice.

Cavity quantum electrodynamics with cold atoms\cite{Optomechanics0,Optomechanics1} offers
an alternative approach to emulating the AA Hamiltonian\cite{NigelCooper,InOpticalLattice}.
Experimental advances motivated new theoretical studies towards the extension of
the AA model. These studies include incorporation of interaction
effects\cite{Shlyapnikov,Interacting2015}, effects of
the ac-drive\cite{ac-driven},
and the dynamics of a
quench\cite{QuenchDynamics,QuenchDynamics1}.

Another recent development in the field of cold atoms
is the possibility to impose Zeeman shifts and
spin-orbit coupling\cite{Spielman,Galitski-Spielman}
by illumination
the condensate with lasers.
%with a pair of
%laser beams.
This raises a question about the extension of the AA
model to incorporate spin-dependent effects.
We address this question in the present paper.

The result of incorporation of the Zeeman splitting, $2\Delta$,
into the AA model is, obviously, two decoupled AA models
for $\uparrow$ and  $\downarrow$ spin projections.
At the transition, $V=2t$,  two delocalized states emerge at
energies $\pm \Delta$.
We will show that incorporation of spin-orbit coupling alone
does not violate the duality and amounts to modification
of the hopping matrix element, while
the eigenstates for
%different
opposite chiralities remain degenerate.

Generalization of the AA model becomes nontrivial
when both, Zeeman splitting and spin-orbit coupling,
are incorporated. In this case, {\em the duality is lifted}.
We studied the interplay of the two spin-dependent
effects numerically.  The results are  interpreted
in the limit of large modulation period, $\beta \ll 1$, when the
semiclassical description and, thus,
the language of  phase-space
trajectories\cite{PhaseSpaceTrajectories}
applies. Nontriviality of interplay of Zeeman splitting
and spin-orbit coupling originates from peculiar structure of the
phase-space trajectories and the evolution of this
structure with energy. In the language of
phase-space trajectories, delocalization transition
corresponds to the connectivity of these trajectories both
in coordinate and in momentum space.

Our main finding is that, in the vicinity of the
delocalization transition, a weak spin-orbit coupling
leads to {\em metallization} in the energy domain
$(-\Delta,\Delta)$. The origin of the effect is
that spin-orbit coupling opens new transport channels.
These channels facilitate the coupling between
disconnected trajectories, thus allowing to avoid tunneling.
In general, we demonstrate that restructuring of the
phase-space at certain energy causes an anomaly in
the localization length at this energy even if the
restructuring takes place in the insulating
regime.

\section{phase-space trajectories and localization length in Aubry-Andr{\'e} model in the semiclassical limit}

\subsection{AA model with long modulation period}

The most transparent scenario of  delocalization transition in the
Aubry-Andr{\'e} model was proposed in Ref. \onlinecite{Suslov}.
This scenario is based on simplification which becomes possible
when the inverse period is small, $\beta \ll 1$,
the fractional part, $\beta_1$, of $\beta^{-1}$ is small,
and all the successive $\beta_i$, the fractional parts of
$\beta_{i-1}^{-1}$, are small.
In this limit the system is characterized
by the hierarchy of periods
\begin{equation}
\label{eqsuslov}
l_1 \sim\frac{1}{\beta},~l_2\sim\frac{1}{\beta\beta_1},~...~,~ l_n \sim\frac{1}{\beta\beta_1\beta_2...\beta_{n-1}}, ~ ...
\end{equation}
With $l_n$
%period
growing exponentially with $n$,
the renormalization-group procedure is applicable.
As a first step, smallness of $\beta$ guarantees
that a given period of the  potential, $V(x)=V\cos(2\pi\beta x)$,
contains many, $\sim \frac{1}{\beta}\left(\frac{V}{t}\right)$, levels. These levels are the eigenfunctions of
the operator, $2t\cos {\hat p} + V(x)$, where
the coordinate, $x=n$,  and the momentum, ${\hat p}=-id/dx$,
are treated as continuous variables.
%where the coordinate, $x=\beta n$,  and the momentum, ${\hat p}=-id/dx$,
%are treated as continuous variables, contains many levels.
%Moreover,

By virtue of the same condition, $\beta \ll 1$,
the levels are discrete, i.e. the overlap, $t^{(1)}$, of the
wave functions in the neighboring periods
is smaller than the level spacing.
If a group of $\beta^{-1}$ sites is viewed as a ``supersite",
which is the essence of the renormalization group transformation,
%$\sim \beta V$.
then this overlap plays the role of
%is viewed as
a first-order hopping integral.
On the other hand, with overlap neglected, the levels
in the neighboring periods are mismatched.
This mismatch, being the result of irrationality of $\beta$,
plays the role of the first-order modulation amplitude, $V^{(1)}$, of the
supersite energies\cite{Suslov}.
As a result, original model Eq. (\ref{eqh0}) with parameters $t$, $V$,
and the period, $\beta^{-1}$, transforms into the same model with
renormalized parameters $t^{(1)}$, $V^{(1)}$, and the period,
$\beta_1^{-1}$. From the fact that renormalized Hamiltonian possesses
duality, the recurrent relation put forward in Ref. \onlinecite{Suslov}
has the form
\begin{equation}  
\label{recurrent}
\left(\frac{V^{(n+1)}}{2t^{(n+1)}}-1\right)
\sim \frac{1}{\beta_{n}} \left(\frac{V^{(n)}}{2t^{(n)}}-1\right)\sim l_{n+1} \left(\frac{V}{2t}-1\right)^n.
\end{equation}
The fact that the critical exponent in the AA model
is equal to $1$ follows immediately from Eq. (\ref{recurrent}).

%Renormalization proceeds until the maximal step, $N$, for which
%the assumption of large number of levels per period, $\frac{1}{\beta_{N}}\left(\frac{V^{(N)}}{t^{(N)}}\right)\gg 1$,
%is violated.

On the physical level, the above procedure captures
how the allowed band at the step, $n$, breaks into
$\frac{1}{\beta_{n+1}}$ allowed bands at the step $(n+1)$ 
(devil's staircase\cite{Azbel1979}).
On the quantitative level, besides the critical exponent,
this description does not answer the basic
question about the energy dependence of the localization
length in the insulating regime. We focus on this
question in our numerical study.

In numerics, the full-developed staircase cannot be captured.
Still,  with two spin-dependent effects incorporated,
the duality of the model gets  violated,  
so that the modification of the localization
length turns out to be highly nontrivial.

%Quantitatively, this is expressed by the condition
%$\oint p(E_n,x)dx =2\pi\hbar n$  are discrete

% The states within different
%periods can be represented as closed contours in the phase space, see Fig. \ref{fpercolation}.
%defined by the equation
%
%The sizes of the contours found from the semiclassical condition
%$\oint p(E_n,x)dx =2\pi\hbar n$  are discrete, but dense, since $\beta \ll 1$.

\subsection{Test of the numerical approach}
\begin{figure}
\includegraphics[scale=0.3]{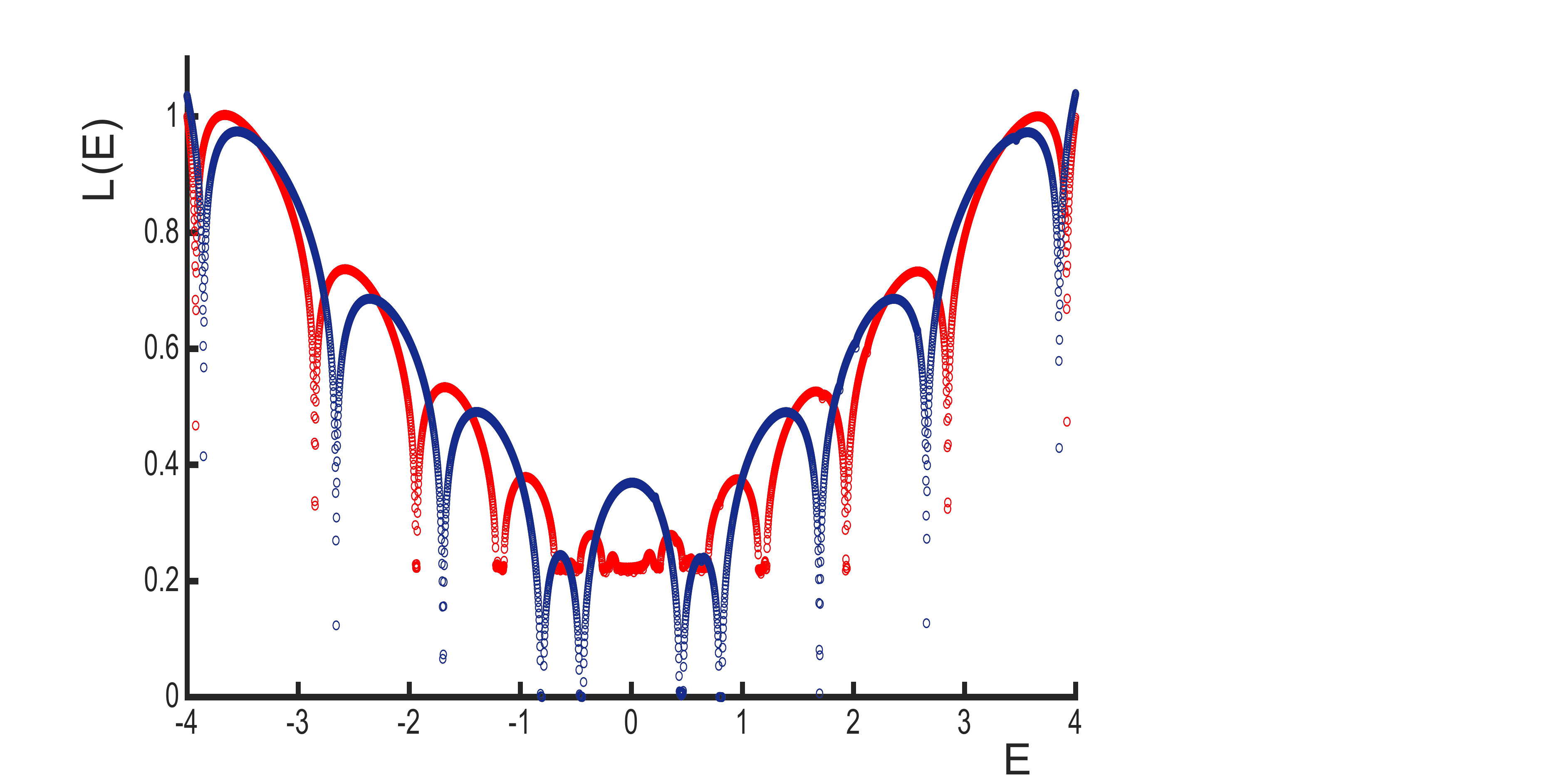}
\caption{(Color online) Energy dependence of the Lyapunov exponent, $L(E)$, calculated numerically
in the insulating regime with $t=0.4V$ for two values of $\beta$:  commensurate $\beta=\frac{1}{10}$ (blue) and incommensurate $\beta=\frac{\sqrt{5}-1}{14}$ (red).  While the values of $\beta$ are close to each other, $L(E)$ for incommensurate potential
exhibits a plateau. The band-structure-induced wiggles are weak within the plateau.
In the semiclassical limit,
the plateau is expected in the interval $|E|<(V-2t)=0.2V$.
Energy is measured in the units of $V/2$. }
\label{foscillations}
\end{figure}

Numerical studies of the localization properties of eigenfunctions in AA model are carried out
either by analysis of the inverse participation ratio, see e. g. Ref. \onlinecite{DasSarmaEdges},
or by the analysis of eigenvalues of the transfer matrix, as in Ref.  \onlinecite{SO+bosons}.
We have adopted the approach suggested in Ref. \onlinecite{NUMERICAL},
which is based on the Thouless formula.\cite{Thouless1972}
The object of interest is the behavior of the Lyapunov exponent, $L(E)$, which is the inverse localization length of the state with energy, $E$.
The details of the computational procedure are presented in Appendix I.
The analysis of the numerical data is complicated by the fact that irrational period of
the potential, $\beta$, is approximated as a rational number in the computational process.
For rational $\beta$, the exponent, $L(E)$, turns to zero within the energy bands, separated
by the gaps. This causes ``wiggles" in $L(E)$ obtained numerically. We call these wiggles, the
band-structure effect. The problems caused by the  band-structure effect become less acute
as the value of $\beta$ is decreased.  This is because the bands of allowed energies become
progressively small. Still, it is important to note that, even at small $\beta$, the
exponent, $L(E)$, calculated numerically, is distinctively different for incommensurate $\beta$.
This is illustrated in Fig. \ref{foscillations}, where the results for $L(E)$ are shown
for two close values of $\beta$, one rational, $\beta =\frac{1}{10}$, and one irrational, $\beta =\frac{\sqrt{5}-1}{14}$.
%We checked this conjecture numerically. The results are shown in Fig. \ref{foscillations}.
%First we see that the numerical procedure described in Appendix I distinguishes
%between the rational and irrational $\beta$.
%Namely, the values $\beta =\frac{1}{10}$ and $\beta =\frac{\sqrt{5}-1}{14}$ are close to each
%other.
For irrational $\beta$, the expansion into continuous fraction has the form: $\beta_1=11$, $\beta_2=3$, $\beta_3 =15$, and $\beta_i=\beta_{i-2}$ for $i>3$.
%The behavior of the Lyapunov exponents as a function of energy are dramatically
%different for these two values.
The main difference between the two $L(E)$ curves is that, for irrational $\beta$,
the $L(E)$-dependence exhibits a continuous plateau in the  insulating regime, $V>2t$. The width of the plateau
is $2(V-2t)$, reflecting the proximity to the transition. On the contrary,
for rational $\beta$, $L(E)$ has two zeros in the plateau region connected by a smooth curve. This indicates that, for  rational $\beta$, the behavior of eigenfunctions is not sensitive to the
transition.
%In the insulating regime, $V>2t$, For irrational $\beta$, the curve $L(E)$ exhibits
%a plateau in the domain $|E|< (V-2t)$.
Outside of the interval $|E|< (V-2t)$ both $L(E)$ curves have similar behavior.
The difference between rational and irrational $\beta$ manifests itself in the fact
that, for irrational $\beta$, the Lyapunov exponents, while exhibiting minima
indicative of underdeveloped band-structure, never turns to zero.

% exhibits a general growth with energy interrupted with minima. This minima, however, never reach zero.
%
% As seen in Fig. \ref{foscillations}, in the insulating regime, $V>2t$,  $L(E)$ with rational $\beta$,  drops
%to zero at numerous energies. This is a manifestation of the presence of a true band-structure. On the contrary, for irrational $\beta$, with $\beta_1=11$, $\beta_2=3$, $\beta_3 =15$, and $\beta_i=\beta_{i-2}$ for $i>3$,  the curve $L(E)$ exhibits
%a plateau in the domain $|E|< (V-2t)$. Outside of this plateau $L(E)$ exhibits a general growth with energy interrupted with minima. This minima, however, never reach zero.

\section{AA model in the semiclassical limit}

\begin{figure}
\includegraphics[scale=0.25]{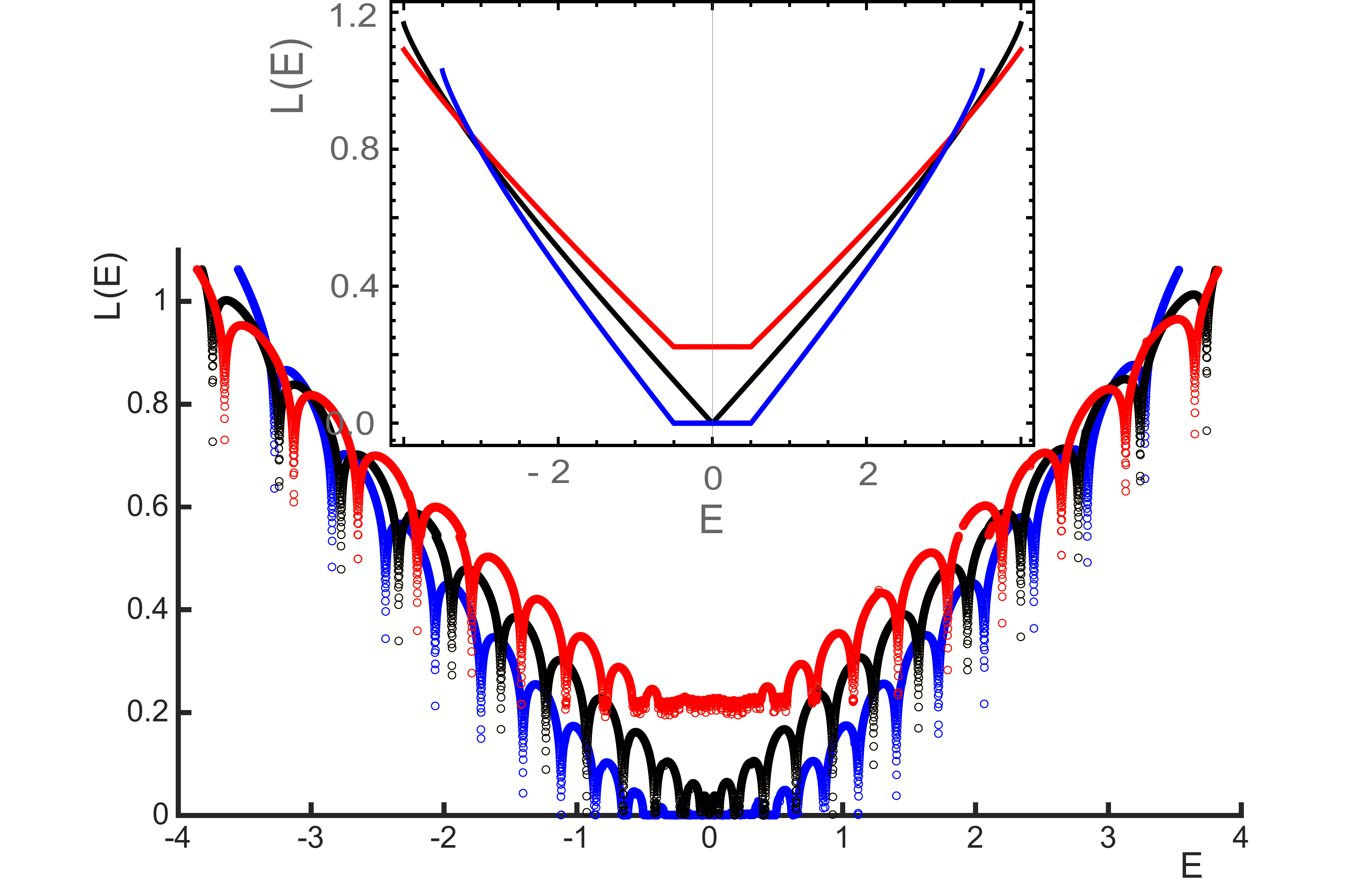}
\caption{(Color online) Manifestation  of the delocalization transition  in the energy dependence of the Lyapunov exponent.
The curves, $L(E)$,
calculated numerically for $\beta=\frac{\sqrt{5}-1}{29}$ and for the same hopping integrals as in Fig. \ref{fpercolation}: $t=0.4V$ (red), $t=0.5V$ (black),
and $t=0.66V$ (blue), are shown. Wiggles in the curves reflect the ``band-structure effect," as in Fig. \ref{foscillations}. Still, the
smooth parts of the curves clearly indicate the transition from insulator to metal.
%In fact,
These smooth parts are described very well by $L(E)$ calculated
analytically from Eqs. (\ref{Lfinal1}) and (\ref{Lfinal2}) and shown in the inset.}
\label{fnumerical}
\end{figure}

As discussed in the previous Section, incommensurability manifests itself as a plateau in $L(E)$
dependence. Upon decreasing $\beta$, the plateau becomes progressively more pronounced. This is because wiggles in $L(E)$ get suppressed. This is illustrated in  Fig. \ref{fnumerical} for
$\beta=\frac{\sqrt{5}-1}{29}$, i.e. two times smaller than in Fig. \ref{foscillations}.
It is important that, despite the presence of wiggles, the evolution of $L(E)$ across the
metal-insulator transition can be clearly traced. The plateau in the insulating regime evolves
into $L(E)\propto |E|$ at the transition. This is followed by a plateau $L(E)=0$ in the metallic regime. The behavior $L(E)\propto |E|$ in the interval $|E|<(V-2t)$ is in accord with general theory\cite{Suslov}.

\begin{figure}
\includegraphics[scale=0.5]{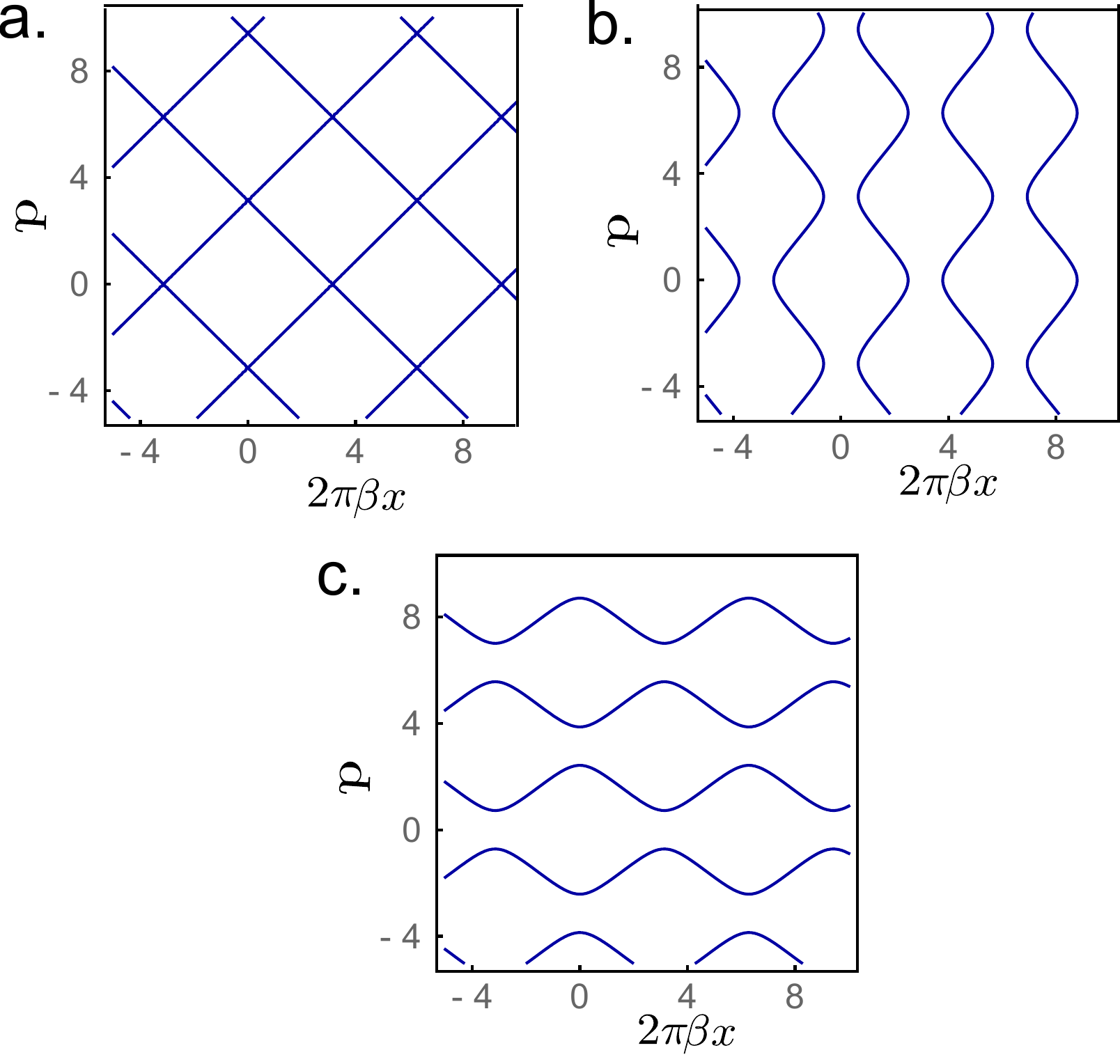}
\caption{(Color online) Phase-space trajectories in the AA model are shown for zero energy and for three values of the hopping integral: $t=0.5V$ (a),
$t=0.4V$ (b), and $t=0.66V$ (c). Trajectories in (b) ``percolate" in the $p$-direction, while the trajectories in (c) percolate in
the $x$-direction. }
\label{fpercolation}
\end{figure}

In this Section we demonstrate that, aside from wiggles, the behavior of $L(E)$ in Fig. \ref{fnumerical}
can be captured {\em quantitatively} within the semiclassical description.
For small $\beta$ the potential changes slowly, which allows to introduce a local dispersion law
at a given $x$. This law has a form
\begin{equation}
\label{eqclassical1}
E(p,x)=V\cos (2\pi\beta x) + 2t \cos p.
\end{equation}
The equation Eq. (\ref{eqclassical1}) defines a system of phase-space trajectories,\cite{PhaseSpaceTrajectories} $E(p,x)=E$.
These trajectories are illustrated in Fig. \ref{fpercolation}.

\begin{figure}
\includegraphics[scale=0.11]{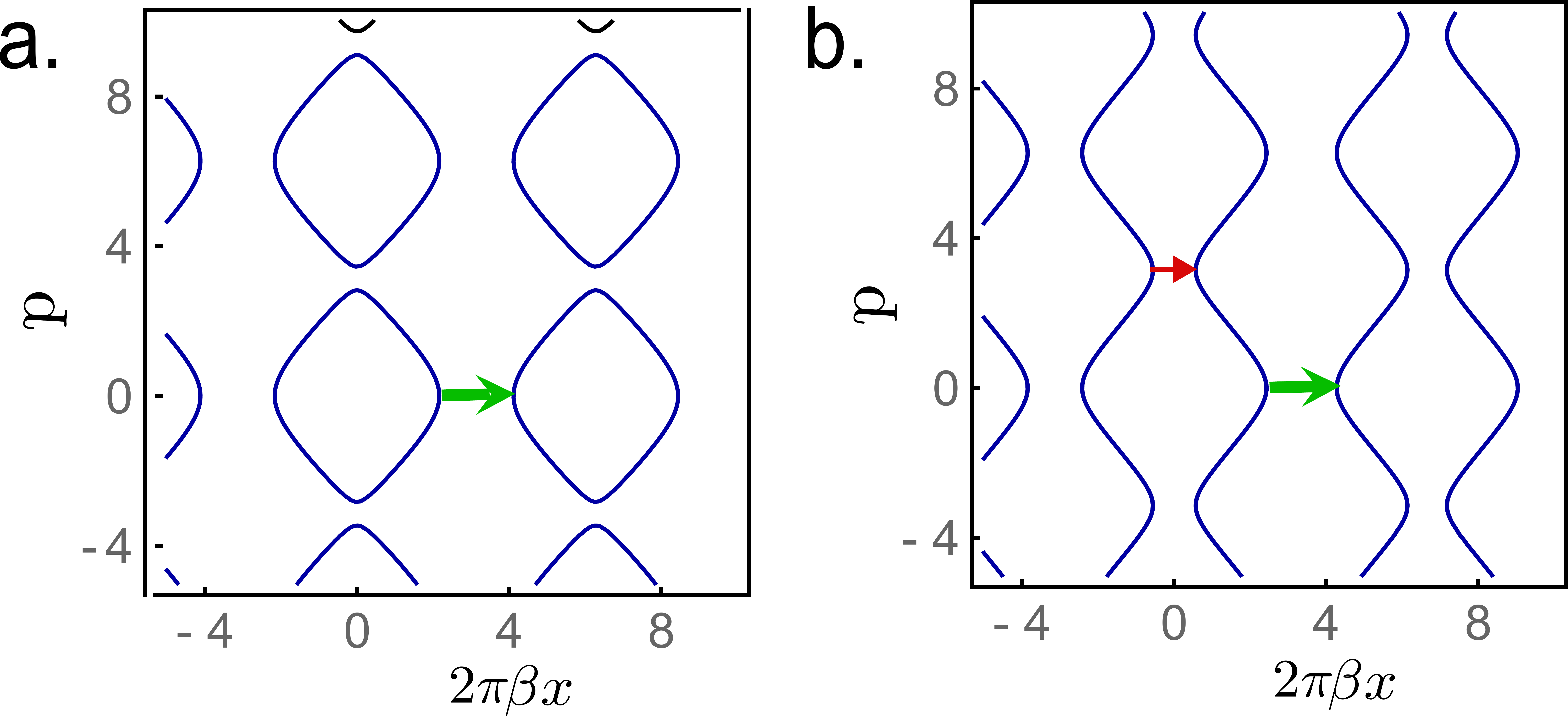}
\caption{(Color online) Illustration of the origin of a plateau in $L(E)$ which develops in the
interval $|E|<(V-2t)$ in the
insulating regime. (a) For energies outside of this interval  the coupling of the phase-space trajectories separated by a period involves
a one-step tunneling at $p=0$, which is the shortest tunneling path. (b) For energies inside the interval this coupling involves a
two-step tunneling at $p=0$ (green arrow) and $p=\pi$ (red arrow). As follows from Eq. \ref{L1expanded}, the logarithm of the tunneling amplitude for the first step is proportional to $(V-2t +E)$, while for the second step this logarithm is proportional to
$(V-2t-E)$. Thus, the product of the amplitudes does not depend on $E$.   }
\label{finsulator}
\end{figure}
In the semiclassical limit, the duality, $p\leftrightarrow 2\pi x$ and $V\leftrightarrow 2t$,
becomes apparent. In the language of  phase-space trajectories, the metallic and the insulating
states correspond to the trajectories continuous in the $x$-direction and  discontinuous in $x$-direction, respectively. Metal-insulator transition at $V=2t$ takes place when the trajectories, corresponding to $E=0$, ``percolate". The energy dependence of the Lyapunov exponent at the transition point is determined by tunnel coupling of the trajectories disconnected in the
$x$-direction. It follows from Eq. (\ref{eqclassical1}) that the tunneling takes place either
along the line ${\text {Re}}~p=0$ or along the line ${\text {Re}}~p=\pi$. These points correspond to the minimal separation of the disconnected trajectories, see Fig. \ref{finsulator}. Using Eq. (\ref{eqclassical1}),   the semiclassical expression for the logarithm of the coupling constant, $\int dx~ {\text {Im}}\hspace{1mm}p$, calculated along ${\text {Re}}~p=0$, can be cast in the form
\begin{equation}
\label{Lfinal1}
{\cal L}_1=\frac{1}{\pi \beta}\int\limits_{0}^{\cosh^{-1} \left(\frac{V+E}{2t}\right)} \hspace{-5mm}dq\frac{q\sinh q}{\sqrt{\big(\frac{V+E}{2t}-\cosh q\big)\big(\frac{V-E}{2t}+\cosh q\big)}}.
\end{equation}
Corresponding expression for tunneling along ${\text {Re}}~p=\pi$ reads

\begin{equation}
\label{Lfinal2}
{\cal L}_2=\frac{1}{\pi \beta}\int\limits_{0}^{\cosh^{-1} \left(\frac{V-E}{2t}\right)} \hspace{-5mm}dq\frac{q\sinh q}{\sqrt{\big(\frac{V+E}{2t}+\cosh q\big)\big(\frac{V-E}{2t}-\cosh q\big)}}.
\end{equation}
The upper limit in Eq. (\ref {Lfinal1}) corresponds to the first bracket in the denominator turning to zero, while the upper limit of Eq. (\ref {Lfinal2}) corresponds to the second bracket in the denominator turning to zero. If the argument in $\cosh^{-1}$ is smaller than $1$, the
corresponding ${\cal L}$ should be set to zero.

At critical value $V=2t$ only  ${\cal L}_1$ is nonzero for $E>0$, while only ${\cal L}_2$ is nonzero for $E<0$. For $E\ll t$ we can expand $\cosh q$ in the first bracket and replace $\cosh q$ by $1$ in the second bracket. Then the integral can be readily evaluated yielding
\begin{equation}
\label{L1expanded}
{\cal L}_1=\frac{1}{\beta}\left(\frac{E}{4t} \right).
\end{equation}
To relate ${\cal L}_1$ to the Lyapunov exponent, we reason as follows.
Tunnel coupling of two trajectories separated in the $x$-direction by $n$ periods is $\exp(-n{\cal L}_1)$. The distance between this trajectories is $x_n=\frac{n}{\beta}$. Expressed via the
 Lyapunov exponent, this coupling is $\exp(-Lx_n)$. Thus, $L$ and ${\cal L}_1$ are related as
 $L=\beta {\cal L}_1$. We then conclude that the semiclassical result Eq. (\ref{L1expanded})
 captures the behavior of the Lyapunov exponent at the transition obtained numerically and shown in Fig. \ref{fnumerical}.

 Consider now the vicinity of the transition $0<(V-2t)\ll t$. In the
  domain $|E|>(V-2t)$ only one of ${\cal L}_1$, ${\cal L}_2$ is nonzero, as it was at the transition.
Then the generalization of Eq. (\ref{L1expanded}), valid for arbitrary sign of $E$, takes the form
\begin{equation}
\label{eqLEoutside}
L(E)=\frac{ |E|+(V-2t)}{4t}.
\end{equation}
In the domain $|E|<(V-2t)$ both ${\cal L}_1$ and ${\cal L}_2$ are nonzero. The Lyapunov exponent is determined by the sum
\begin{equation}
 \label{2contribution}
 L(E)=\beta \left(  {\cal L}_1 + {\cal L}_2\right).
 \end{equation}
It is easy to see that the energy drops out from this sum, so that
\begin{equation}
\label{eqplateu}
L(E)=\frac{(V-2t)}{2t}
\end{equation}
in this domain. The results Eq. (\ref{eqLEoutside}) and Eq. (\ref{eqplateu}) are plotted  in Fig. \ref{fnumerical}, inset. We see that they completely agree with numerical results shown in the same figure. The expression for $L(0)$ is in accord with critical exponent of the AA model being equal to $1$.\cite{Suslov}

We note that the simulation of the $L(E)$-dependence
was previously carried out in Ref. \onlinecite{NUMERICAL}. To suppress the band-structure effects the on-site energies were chosen in the form $V\cos(2\pi\beta |n|^{\nu})$, with $\nu=0.7$, so
that the results did not depend on whether or not $\beta$ is irrational. Numerical results in Ref. \onlinecite{NUMERICAL} are
quite similar to those shown in Fig. \ref{fnumerical}, inset. However, the authors did not have an explanation for the plateau.

It is instructive to illustrate the $L(E)$-behavior in the AA model with the help of Fig. \ref{finsulator}. Coupling between two phase-space trajectories separated by a period, $1/\beta$, requires tunneling. For $|E|>(V-2t)$, the geometry of
the trajectories is such that this tunneling is a one-step process,
see Fig. \ref{finsulator}a. By contrast, for $|E|<(V-2t)$ the
geometry of the trajectories is different, so that one-step tunneling is insufficient for the transport along $x$. Rather, the coupling is a product of the amplitudes of tunneling
at $p=0$ and $p=\pi$. Upon the change of energy, one amplitude
grows, while the other amplitude drops off, so that their product remains constant.  Note finally, that the linear behavior of $L(E)$ given by Eq. (\ref{eqLEoutside}) also applies for $V<2t$, outside the metallic domain, $|E|<(2t-V)$.

\section{Delocalization in the presence of the Zeeman splitting: effect of
a weak spin-orbit coupling}

Zeeman splitting is incorporated into the AA model by adding the term $\Delta\sigma$ to the on-site energies,
where $\sigma$ takes the values $\pm 1$.
Presence of spin-orbit coupling
allows a spin-flip process upon hopping to the neighboring site\cite{japanese}. For hopping, say, to the right, we denote the corresponding hopping amplitude
with $it_1$. Then, for hopping to the left, this amplitude is $-it_1$.  With Zeeman splitting
 and spin-orbit coupling included, the Hamiltonian Eq.~(\ref{eqh0}) takes the form
\begin{multline}
\label{eqh1}
\hat{H}=-t\sum\limits_{n,\sigma} \left(c_{n,\sigma}^{\dagger}c_{n+1,\sigma}+ c_{n+1,\sigma}^{\dagger}c_{n,\sigma}\right) \\
-it_1\sum\limits_{n,\sigma\neq\sigma^{'}} \left(c_{n,\sigma}^{\dagger}c_{n+1,\sigma^{'}}- c_{n+1,\sigma^{'}}^{\dagger}c_{n,\sigma}\right)\\
+\sum\limits_{n,\sigma}\Big[V\cos(2 \pi \beta n)+\Delta \sigma\Big]c_{n,\sigma}^{\dagger}c_{n,\sigma}.
\end{multline}
In the semiclassical limit, the Hamiltonian Eq. (\ref{eqh1}) defines two branches of the spectrum and, correspondingly, two types of the phase-space trajectories. They are described by the equations
\begin{equation}
\label{eqclassical2}
V\cos (2\pi\beta x) + 2t \cos p \pm \Big[ \Delta^2 + 4t_1^2 \sin^2 p \Big]^{1/2}=E.
\end{equation}
It is easy to see that, without Zeeman splitting, the effect of spin-orbit coupling
amounts to the replacement of $t$ by $\pm  \left(t^2+t_1^2\right)^{1/2}$. This observation
is, actually, general, i.e. it is valid not only in the semiclassical limit. As demonstrated
in Appendix II, it can be derived rigorously from the Hamiltonian Eq.  (\ref{eqh1}).
Transformation $t \rightarrow \pm \left(t^2+t_1^2\right)^{1/2}$ is accompanied by the shift of momentum. Equally, it is obvious that
without spin-orbit coupling, the eigenstates of the Hamiltonian
Eq. (\ref{eqh1}) are the same as in Eq. (\ref{eqh0}) with eigenvalues shifted by $\pm \Delta$.

\begin{figure}
\includegraphics[scale=0.24]{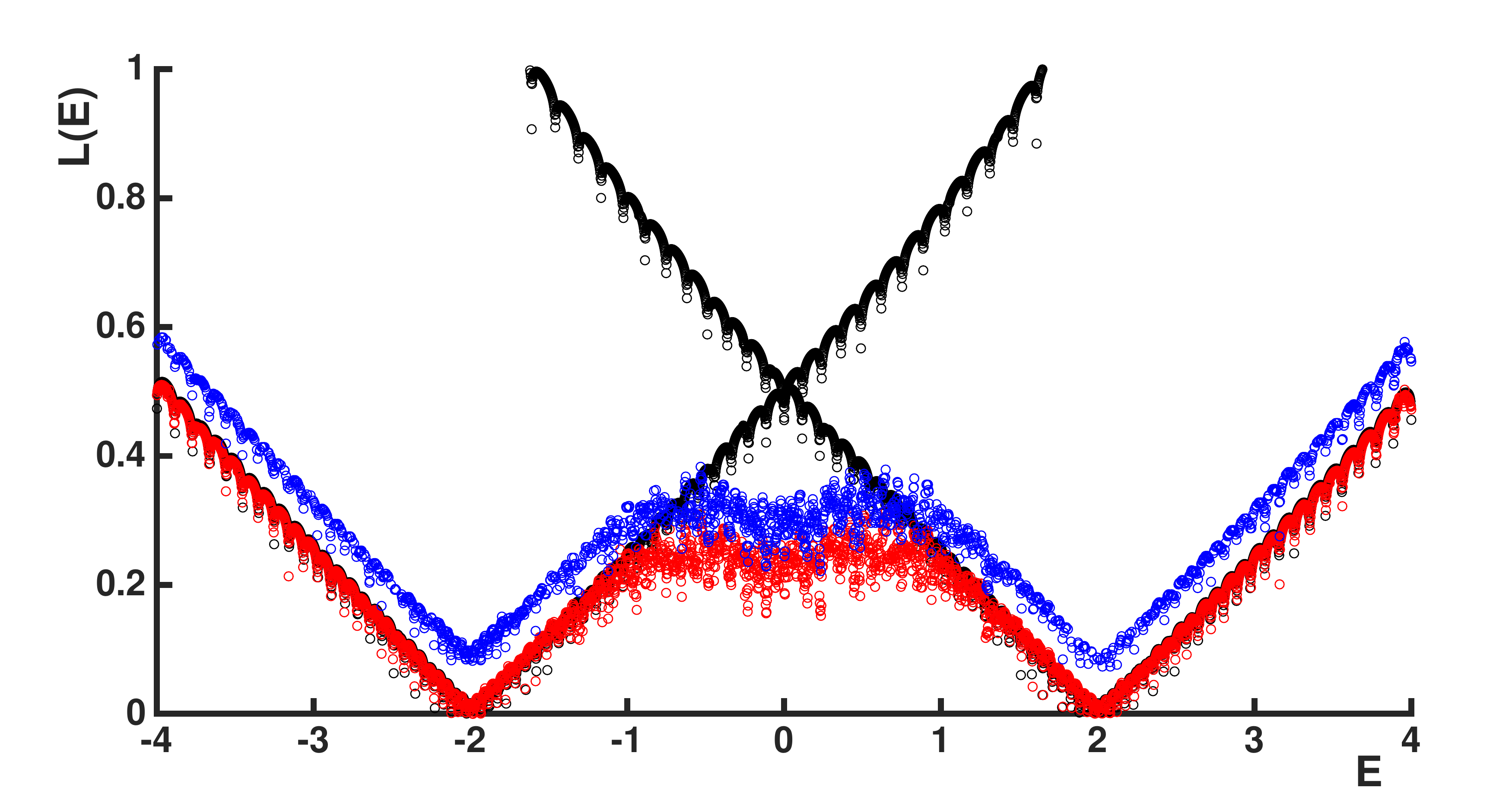}
\caption{(Color online) Illustration of the effect of spin-orbit coupling
on the localization in the AA model with Zeeman splitting.
All the curves show $L(E)$ calculated for the critical value $t=0.5V$.
Zeeman splitting is chosen to be $\Delta=V$.
Black lines are $L(E)$ calculated for $t_1=0$.
They correspond to the metal-insulator
transitions at energies $E=\pm \Delta$.
Incorporation of a weak spin-orbit coupling, $t_1=0.05V$, suppresses the
localization in the domain $-\Delta <E<\Delta$. Blue and red numerical curves are for two eigenvalues of the transmission matrix (see the text).
For clarity, the calculation was performed for large period,
$\beta=\frac{\sqrt{5}-1}{97}$. }
\label{ft1so}
\end{figure}

Our prime finding is that the interplay of the two spin-dependent
processes has a dramatic effect on the localization properties of
the eigenstates. More specifically, very small spin-orbit coupling
leads to a strong suppression of the localization. This is illustrated in Fig. \ref{ft1so}.
The dependence, $L(E)$, in this
figure was calculated for parameters $V$ and $t$ corresponding to
the criticality, so that for, $t_1=0$, we obtained two $V$-shaped
curves centered at $E=\pm \Delta$. After a small $t_1=0.05V$ was
included, the behavior of $L(E)$ near $\pm \Delta$ did not change.
However, $L(E)$ dropped down significantly in a wide domain of  intermediate energies $|E|<V$.

The physics underlying this stark suppression of localization is the following.
Small $t_1$ opens new channels of coupling between the
trajectories corresponding to a given spin. Obviously, for $t_1=0$,
{\em all} eigenstates corresponding to the branches Eq. (\ref{eqclassical2}) are
orthogonal to each other. With finite $t_1$, the eigenstates corresponding to a {\em given momentum} are orthogonal to each other. However, the eigenstates corresponding to different momenta
have a finite overlap. Below we confirm this statement by a direct
calculation.

The analytical forms of the eigenfunctions corresponding to $+$ and
$-$ branches are the following:
\begin{equation}
\label{plusmatrix}
\Psi_p^{+}=\begin{pmatrix}
\varphi_{1p}^{+} \\ \\ \varphi_{2p}^{+}   \end{pmatrix}=\frac{1}{2^{1/2}D_p^{1/4}(D_p-\Delta)^{1/2}}\begin{pmatrix}
-2i t_1 \sin p \\ \\ \Delta- D_p\end{pmatrix},
\end{equation}

\begin{equation}
\label{minusmatrix}
\Psi_p^{-}\hspace{-1mm}=\hspace{-5mm}\quad\begin{pmatrix}
\varphi_{1p}^{-} \\ \\ \varphi_{2p}^{-}   \end{pmatrix}=\frac{1}{2^{1/2}D_{p}^{1/4}(D_{p}+\Delta)^{1/2}}\begin{pmatrix}
-2i t_1 \sin p \\ \\ \Delta+ D_{p}\end{pmatrix},
\end{equation}
where $D_p$ is defined as
\begin{equation}
\label{Dp}
D_p=\left( \Delta^2 + 4 t_1^2 \sin^2 p \right)^{1/2}.
\end{equation}
Using Eqs. (\ref{plusmatrix}) and (\ref{minusmatrix}), we calculate the
scalar product of $+$ and $-$ eigenfunctions with different momenta
and obtain
%$\langle\Psi_{p+\frac{q}{2}}^{-}|\Psi_{p-\frac{q}{2}}^{+}\rangle$ can be cast in the form
\begin{widetext}
\begin{multline}
\label{scalarproduct}
\langle\Psi_{p+\frac{q}{2}}^{-}|\Psi_{p-\frac{q}{2}}^{+}\rangle= -\frac{4 t_1^2\Delta\sin \frac{q}{2} \cos p }{ (D_{p+\frac{q}{2}}D_{p-\frac{q}{2}})^{1/4}[(D_{p+\frac{q}{2}}+\Delta)(D_{p-\frac{q}{2}}-\Delta)]^{1/2}}
\\
\times \Bigg[\frac{\Delta \sin \frac{q}{2} \cos p }{\Delta ^2 +4t_1^2 \sin(p+\frac{q}{2})\sin(p-\frac{q}{2})+D_{p+\frac{q}{2}}D_{p-\frac{q}{2}}}
+ 2\frac{  \cos \frac{q}{2} \sin p }{D_{p+\frac{q}{2}}+D_{p-\frac{q}{2}}} \Bigg].
\end{multline}
\end{widetext}
It is easy to see that this product is zero when either $\Delta=0$, $t_1=0$, or the transferred  momentum, $q$, is equal to zero.
\begin{figure}
\includegraphics[scale=0.45]{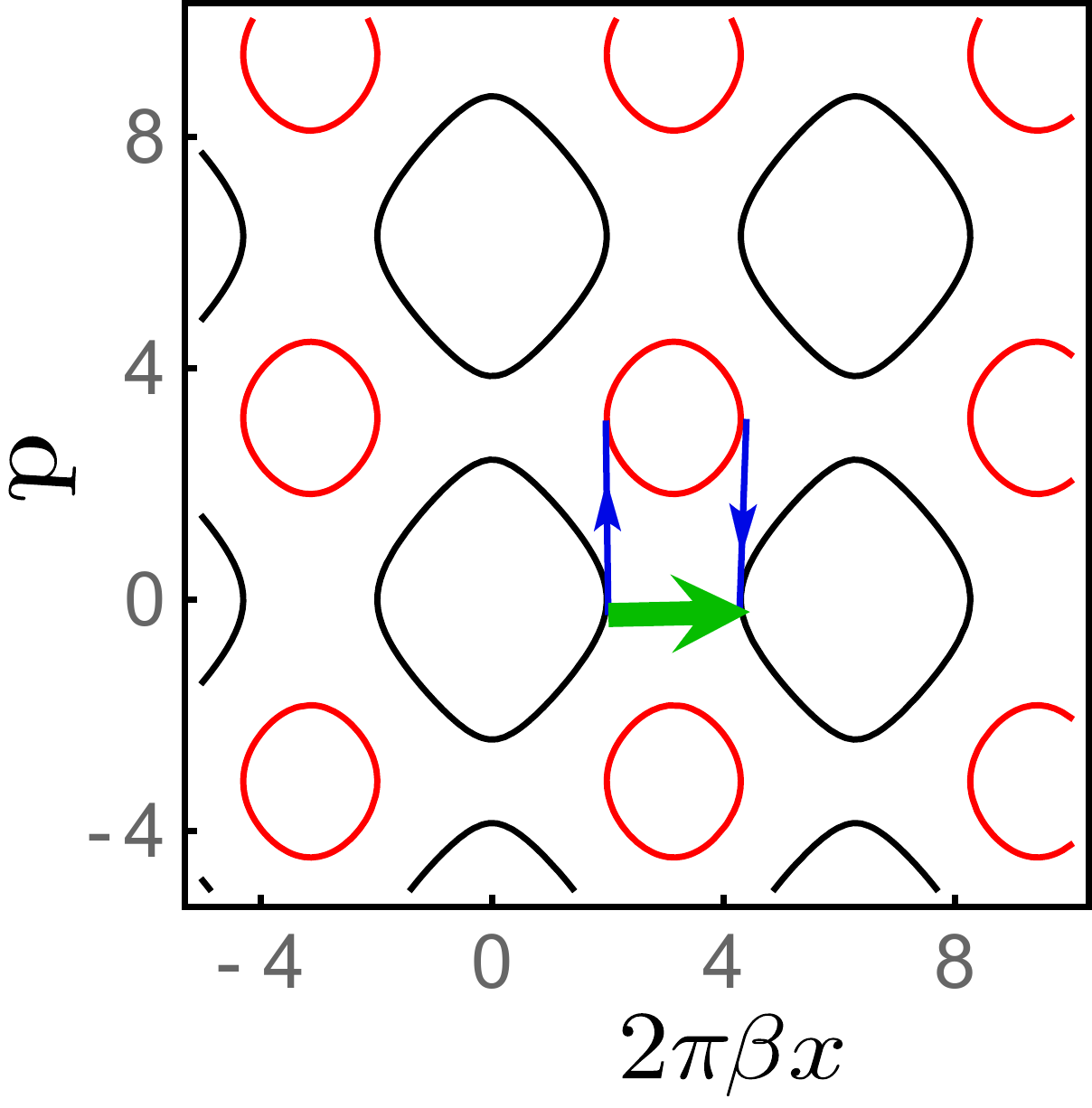}
\caption{(Color online) Schematic illustration of the spin-orbit facilitated delocalization.
The phase-space trajectories for a given energy $E <\Delta$  include the contours corresponding to two different branches, see Eq. (\ref{eqclassical2}). They are shown with black and red lines. Coupling  between two black
contours is determined either by tunneling (green arrow) or by two virtual transitions from black
to red and back (blue arrows). These transitions are accompanied by change of the
momentum by $\pi$, so that the corresponding spinors are not orthogonal, see
Eq. (\ref{scalarproduct}). Moreover, the amplitudes of these transitions are proportional to $t_1$
but do not contain tunneling exponent. Thus, already at  very small $t_1$, these transitions dominate the coupling. As a result,
$L(E)$ drops dramatically, as shown in Fig. \ref{ft1so}. The contours in the figure are shown for parameters: $\Delta=V$, $t=0.5V$ (as in Fig. \ref{ft1so}), and for $E=0.5V$. }
\label{fsoclassical}
\end{figure}

Now we can explain the behavior of $L(E)$ in Fig. \ref{ft1so}.
The phase-space trajectories corresponding to intermediate energies
are shown in Fig. \ref{fsoclassical}. Black contours correspond to $+$ branch, while red contours correspond to $-$ branch. Direct
coupling between two black contours requires tunneling shown with green arrow. Note, however, that the coupling can be realized by a
two-step process via an intermediate red contour: first the virtual
transition from black to red, shown by a left blue arrow, and then the transition from red to shifted black contour, shown by a right
blue arrow. It can be easily shown that the momentum transfer in
both virtual transitions is $\pi$, so that the blue lines are vertical.

\begin{figure}
\includegraphics[scale=0.24]{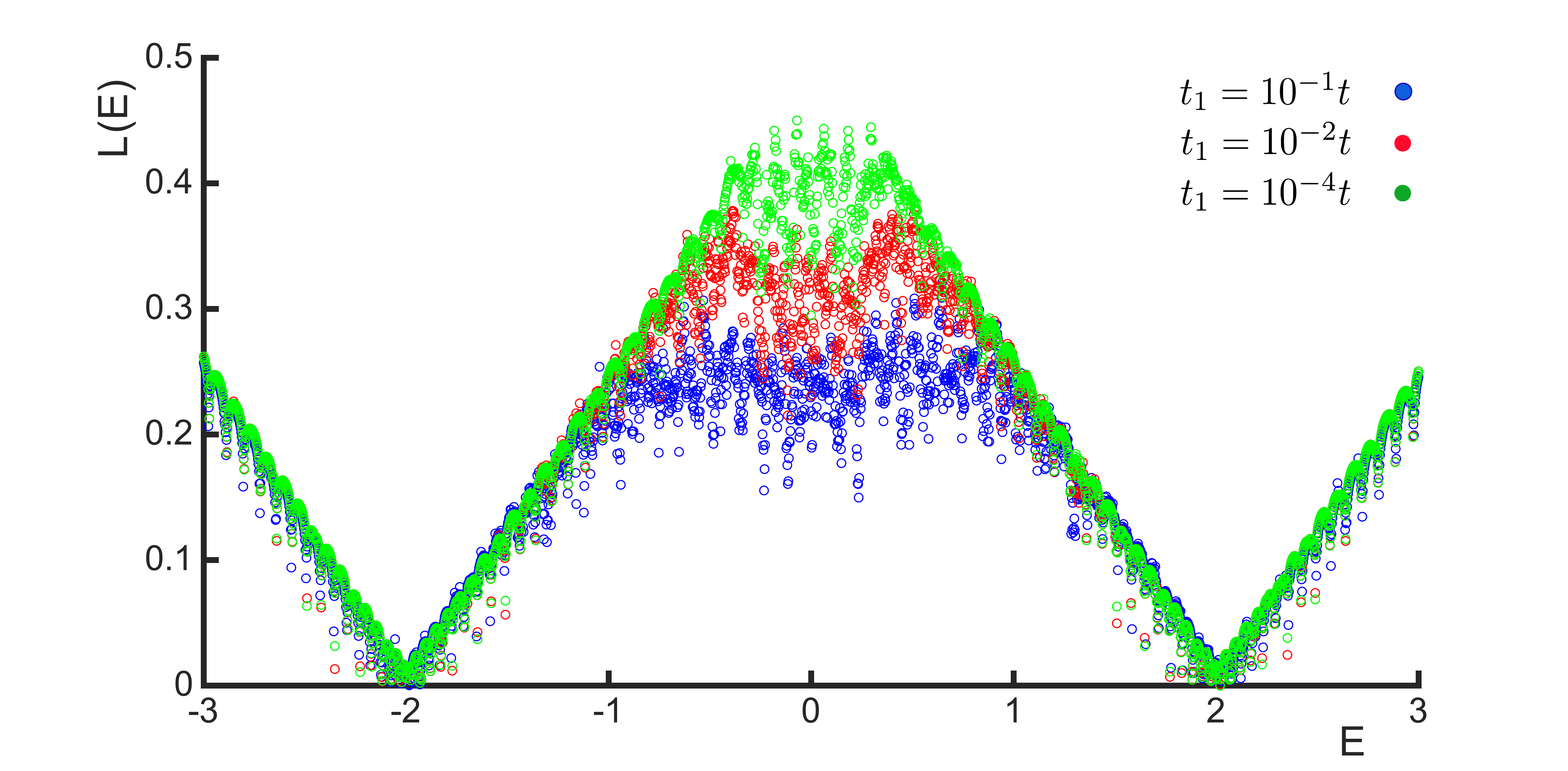}
\caption{(Color online) Scaling of the plateau in $L(E)$ with spin-orbit
strength, $t_1$. The plateau height calculated for three $t_1$-values drops,
upon increasing $t_1$, as $|\ln \left(t_1/t\right)|$. Parameters $\Delta$, $t$ and the period, $\beta$, are the same as in Fig.
 \ref {ft1so}.}
 \label{ft1periodic}
\end{figure}

For small $t_1$, the amplitude of the two-step process
is small $\propto t_1^2$. On the other hand, it does not contain
the tunneling exponent. Thus, this process dominates the Lyapunov exponent when $L(E)$ calculated for direct tunneling is bigger than
$|\ln t_1^2|$. We have checked this prediction numerically. The results are shown in Fig. \ref{ft1periodic}. It can be seen that the plateau in $L(E)$ at intermediate energies indeed scales with
$|\ln t_1^2|$.

To conclude the Section, we demonstrated that for energies, at which the phase-space trajectories corresponding to both branches coexist, a particle can avoid tunneling by ``bouncing" between the states of
different branches. It should be emphasized that this effect is specific only for tight-binding model in which the bandwidth is
limited.

\section{Delocalization due to spin-orbit coupling alone}

In previous Section we assumed that the amplitude, $t_1$, of
hopping with spin-flip constitutes a small correction to the spin-conserving hopping amplitude, $t$.
In the present Section we show that interplay of Zeeman splitting
and spin-orbit coupling {\em alone}, without direct hopping, can result in nontrivial
effects in localization properties of the AA model.

\begin{figure}
\includegraphics[scale=0.24]{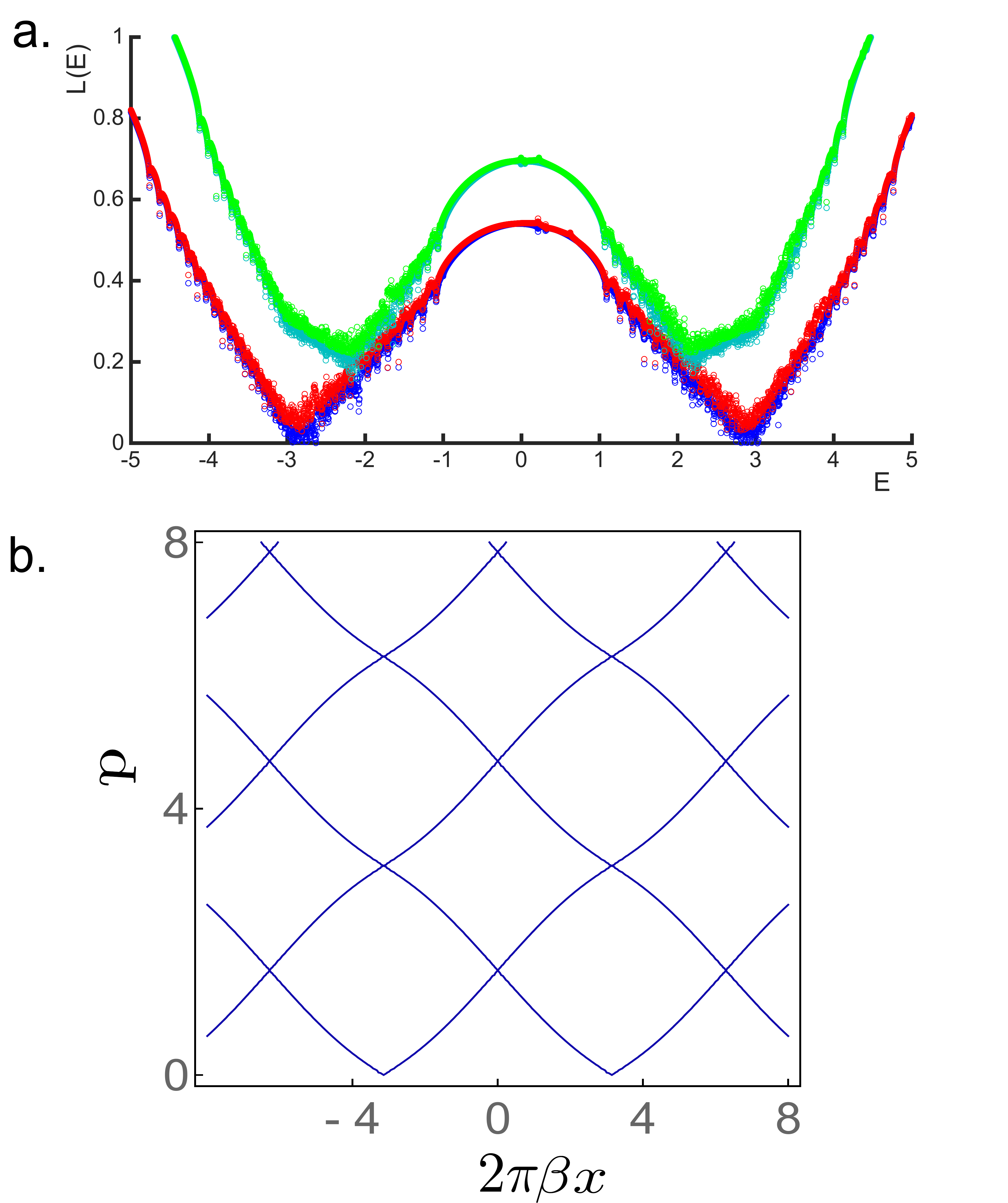}
\caption{(Color online) Illustration of delocalization transition
at zero direct hopping $(t=0)$. The transfer between the sites is
exclusively due to finite spin-orbit coupling. (a): the dependencies $L(E)$,
calculated for $\Delta=2V$ and for two values of $t_1$ are shown: (red) $t_1= 1.72V$
and (green) $t_1=1.22V$. For the first value, the condition Eq. (\ref{t1critical})
is satisfied, so that the phase-space trajectories for $E=E_c=3V$, shown in (b), percolate.
For the second value, $L(E)$ exhibits the plateau-like behavior near the minimum. The underlying
reason for this is that the transport between disconnected phase-space trajectories involves two types
of tunneling similar to Fig. \ref{finsulator}b.
Energy in the figure is measured in the units of $V$. }
\label{ft0smallV}
\end{figure}

Upon setting $t=0$ in Eq. (\ref{eqclassical2}), the equations for the phase-space trajectories assume the form
\begin{equation}
\label{eqclassical3}
\pm \Big[ \Delta^2 + 4t_1^2 \sin^2 p \Big]^{1/2}-E=V\cos (2\pi\beta x).
\end{equation}
Although the duality between $x$ and $p$ is absent, both branches still exhibit the delocalization transition for certain relation between $t_1$, $\Delta$, and $V$. To find this relation and the energy position of the delocalized state, we reason as follows.
Upon changing $p$ from $0$ to $\pi/2$, the combination in the left-hand side of Eq. (\ref{eqclassical3})  %$\left(\Delta^2+4t_1^2\sin^2 p\right)^{1/2}-E$
changes from $\left(\Delta -E\right)$
to $\left[\left(\Delta^2+4t_1^2\right)^{1/2}-E\right]$, while the combination in the right-hand side changes from $-V$ to $V$ upon changing $x$. To achieve percolation of phase-space trajectories, one should
require that these intervals of change coincide. This leads to the
conditions
\begin{eqnarray}
\label{t0percolation}
 \Delta -E=-V,\\
\left(\Delta^2+4t_1^2\right)^{1/2}-E=V.
\end{eqnarray}
Upon solving the above system, we find the critical value of $t_1$ and percolation energy, $E^c$,
\begin{equation}
\label{t1critical}
\frac{t_1^c}{V}=\Bigg[\frac{|\Delta|}{V}+1\Bigg]^{1/2},~~~
\pm \frac{E^c}{V}=\frac{|\Delta|}{V} +1.
\end{equation}
The Lyapunov exponent, $L(E)$, calculated for critical value, $t_1=t_1^c$, and for one value of $t_1$ below the transition  are shown in Fig. \ref{ft0smallV}a. We see that quantum delocalization
indeed takes place at critical $t_1$.
As illustrated in Fig. \ref{ft0smallV}b, the phase-space trajectories at
$t_1=t_1^c$ and $E=E^c$ are not perfect squares. This is the reflection of the absence of $x$-$p$ duality. This duality is respected only near the crossing points, like $(x,p)=(0,\pi/2)$,
but it is these points that are responsible for transport.

\begin{figure}
\includegraphics[scale=0.34]{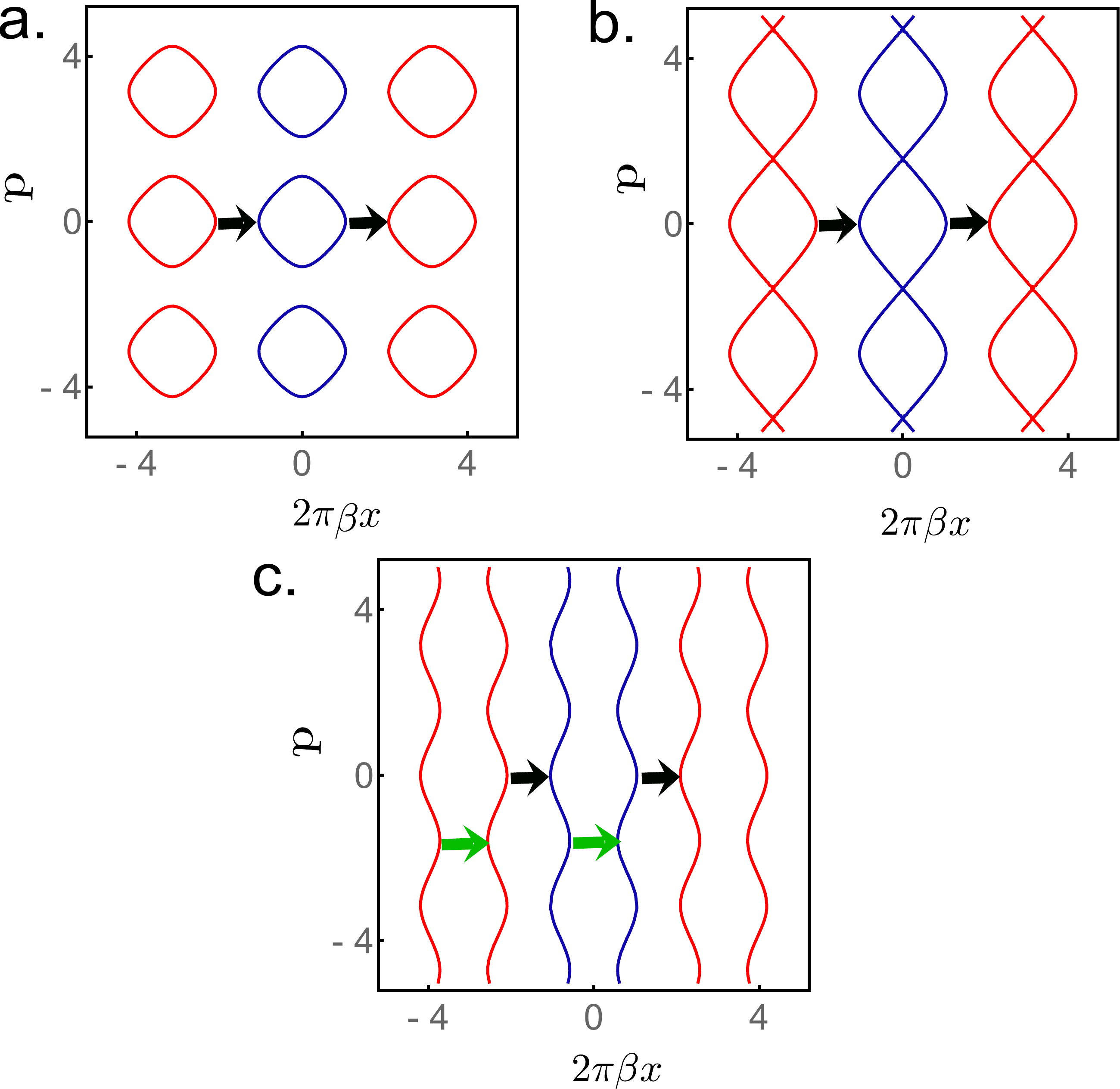}
\caption{(Color online) Phase-space trajectories at zero energy and in the absence of direct
 hopping are shown for parameters $t_1=0.433V$ (a), $t_1=0.333V$ (b), $t_1=0.485V$ (c). Zeeman splitting is chosen to be $\Delta=0.5V$ in all three plots. }
\label{ftobigVclassical}
\end{figure}

Energies $\pm E^c$ correspond to delocalization within individual
branches. Most nontrivial scenario emerges when both branches are involved in transport. We will demonstrate that, for a certain relation between $t_1$, $V$, and $\Delta$ there is an anomaly in the behavior of the localization length with energy {\em inside
the insulator regime}.  This relation is established from the
condition that the energy distance between the branches is equal
to $2V$ and has the form
%In the absence of the Zeeman splitting, the delocalization transition takes
%place at $V=2t_1$ at at critical energy $E=0$. We studied how small but finite
%$\Delta$ affects this transition.
\begin{equation}
\label{Ec}
\frac{4\left({\tilde t}_1^c\right)^2}{V^2}+\frac{\Delta^2}{V^2}=1.
\end{equation}
Firstly, at this $t_1$, the phase-space trajectories corresponding to both branches coexist. They are shown by blue and red lines in Fig. \ref{ftobigVclassical}. The value $t_1={\tilde t}_1^c$ is distinguished by the fact that
the restructuring of the phase-space trajectories corresponding to $E=0$ takes place at this $t_1$. Note that the restructuring
at $t_1={\tilde t}_1^c$ {\em does not involve percolation}, as it is illustrated in Fig. \ref{ftobigVclassical}.

The restructuring of the phase-space trajectories affects the transport for the following reason.
As seen in Fig. \ref{ftobigVclassical}b,  for
$t_1 >{\tilde t}_1^c$, the transport is exclusively due to tunneling between blue and red trajectories. On the contrary, for  $t_1 <{\tilde t}_1^c$ the
transport requires both: inter-branch tunneling between blue and red trajectories as well
as intra-branch tunneling between blue trajectories and between  red trajectories.
This is because for $t_1 <{\tilde t}_1^c$ additional classically forbidden regions appear,
see Fig. \ref{ftobigVclassical}c.
%Instead, the trajectories
%corresponding to {\em different} branches have a minimal
%separation in the $x$-direction.
%It we fix $E=0$ and vary
%$t_1$ around ${\tilde t}_1^c$, then for $t_1>{\tilde t}_1^c$,
%the gap opens in $p$-direction for both types of the classical
%trajectories. The trajectories of each given type become {\em disconnected}
%in $p$-direction.  For $t_1<{\tilde t}_1^c$ the trajectories of each given type
%are connected in $p$-direction. Traveling over the potential landscape
%requires {\em two} types of tunneling: between the branches and within
%as given branch (for $t_1<{\tilde t}_1^c$).

\begin{figure}
\includegraphics[scale=0.18]{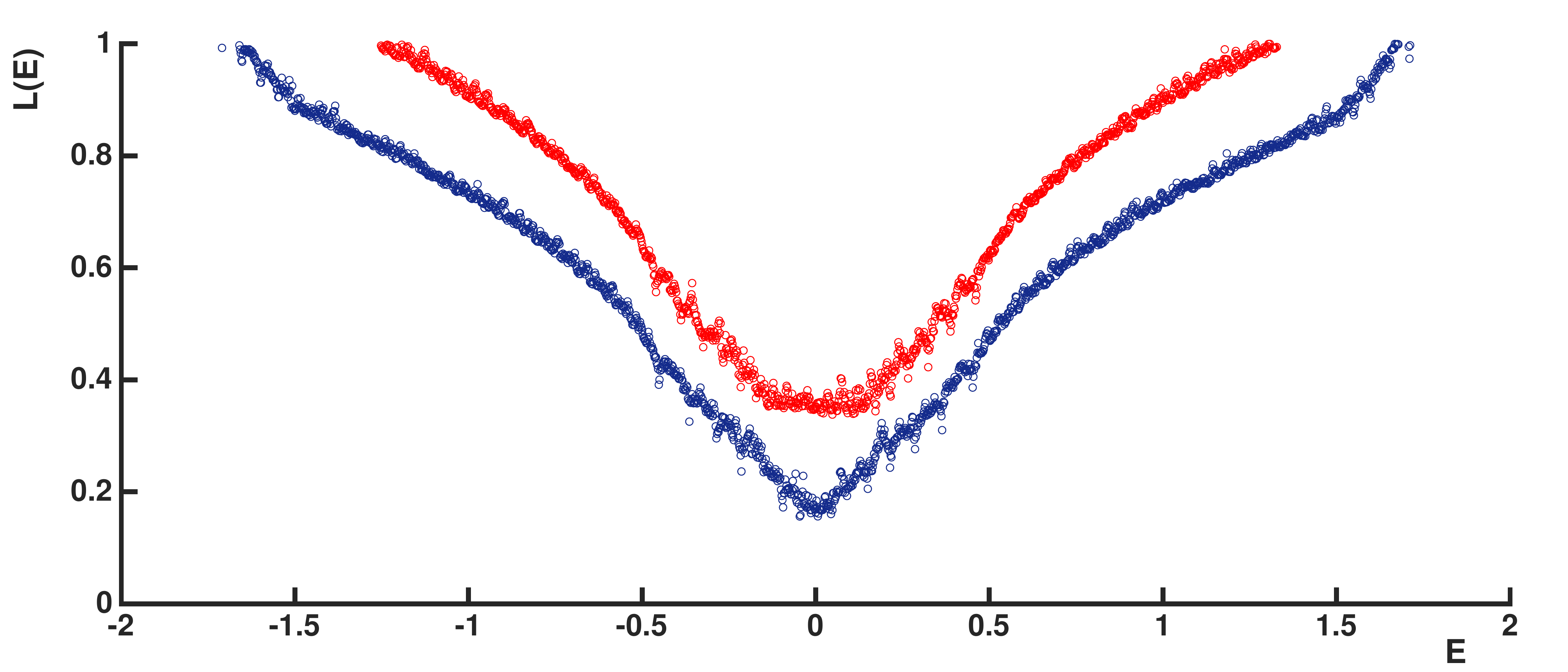}
\caption{(Color online) The dependencies, $L(E)$, are shown for $t=0$, $\Delta=0.5V$ and two  values of $t_1$: (blue) $t_1={\tilde t}_1^c=0.433V$ and (red) $t_1=0.35V$. The blue curve exhibits a $V$-shaped behavior which reflects the restructuring of the phase-space
trajectories. The red curve exhibits a plateau around $E=0$. This plateau has the similar origin as the plateau in Fig. \ref{fnumerical}. Energy is measured in the units of $V$.   }
\label{ft0bigVnumerical}
\end{figure}
Restructuring of the trajectories at $t_1={\tilde t}_1^c$
leads to the anomaly in the energy dependence of the Lyapunov exponent. Namely, $L(E)$ exhibits a $V$-shape behavior, as shown in Fig. \ref{ft0bigVnumerical}. The origin of this behavior is the
following.  The minimal
value, $L(0)$, is determined by the inter-branch tunneling.
%%limited by tunneling between different branches.
%The origin of the $V$-shape behavior is the restructuring of the
%phase-space trajectories with energy.
%%is the restructuring of the
%%classical trajectories which takes place at $E=0$ and at $t_1={\tilde t}_1^c$.
For positive $E$, transport requires additional tunneling between the red
trajectories.
For negative $E$, transport requires additional tunneling between the blue
trajectories. This additional tunneling takes place at $p=\pi/2$.
The ``price" of additional tunneling is proportional to $E$, as we
have established above, see Eq. (\ref{L1expanded}).
Thus, the behavior of $L(E)$  at small $E$ has the form
\begin{equation}
\label{singular}
\left(L(E)-L(0)\right)\propto |E|.
\end{equation}
Note that the tunneling between
red and blue trajectories is ``forbidden", in the sense,
that the initial and final states are both at $p=0$. Thus, the
corresponding spinors are orthogonal to each other.
The reason why this tunneling still takes place is
the uncertainty in the momentum, $\delta p$,
of a tunneling particle. This uncertainty can
be viewed as a momentum transfer in the course of
tunneling. Then the overlap integral
Eq. (\ref{scalarproduct}) can be estimated
as $\delta p/\Delta$. The uncertainty is set
by the discreteness of the AA model, i.e. by
the fact that the coordinate in Eq. (\ref{eqclassical3})
takes integer values. This yields, $\delta p \sim \beta^{-1}$.

For $t_1$ slightly smaller than  ${\tilde t}_1^c$,
a plateau in $L(E)$ develops in the
vicinity of $E=0$. The origin of this plateau
is absolutely similar to the origin of the
plateau around zero energy for $t_1=0$
and finite $t$ slightly smaller than $0.5V$.

%We emphasize that, without spin-orbit coupling,
%we had a $V$-shaped behavior of $L(E)$ at the
%metal-insulator transition, when the localization
%radius diverges. For $t=0$ and finite $t_1$ we
%find the $V$-shaped behavior in purely insulating
%regime. Singularity in $L(E)$ develops,
%while the localization radius remains finite.

%\begin{figure}
%\includegraphics[scale=0.5]{t0classical.pdf}
%\caption{(Color online) $ t_1^2=0.75V^2$ }
%\label{ft0classical}
%\end{figure}

\section{Conclusion}
%he states corresponding to different semiclassical branches are orthogonal to each other,
%so that the ``horizontal" tunneling is forbidden. However, tunneling between the states with
%small momentum transfer is allowed. The corresponding  tunneling length is longer, so that this tunneling is less probable. Thus, there is an optimal momentum transfer for which the product
%of tunneling probability and the square of the overlap integral of initial and final spinors
%is maximal.
\begin{figure}
\includegraphics[scale=0.2]{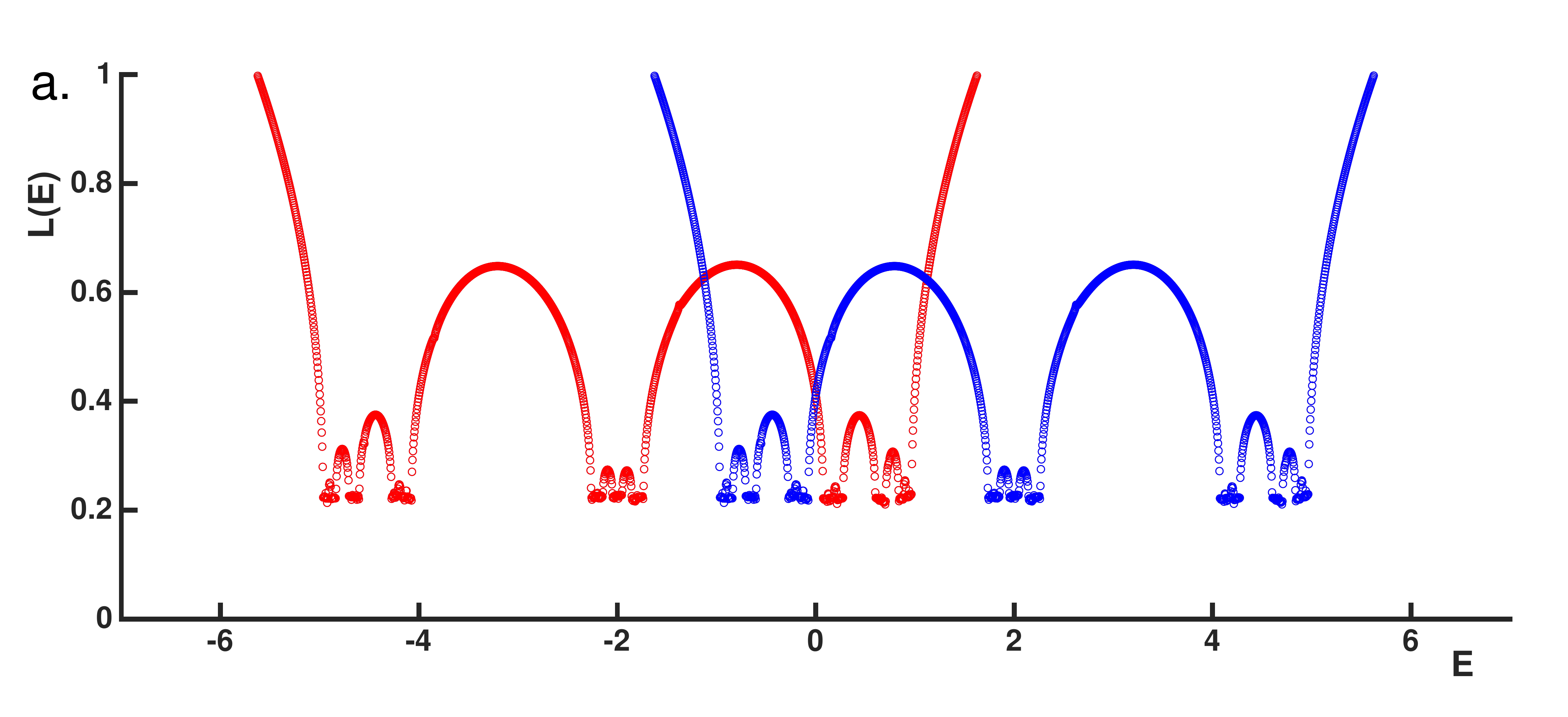}
\includegraphics[scale=0.3]{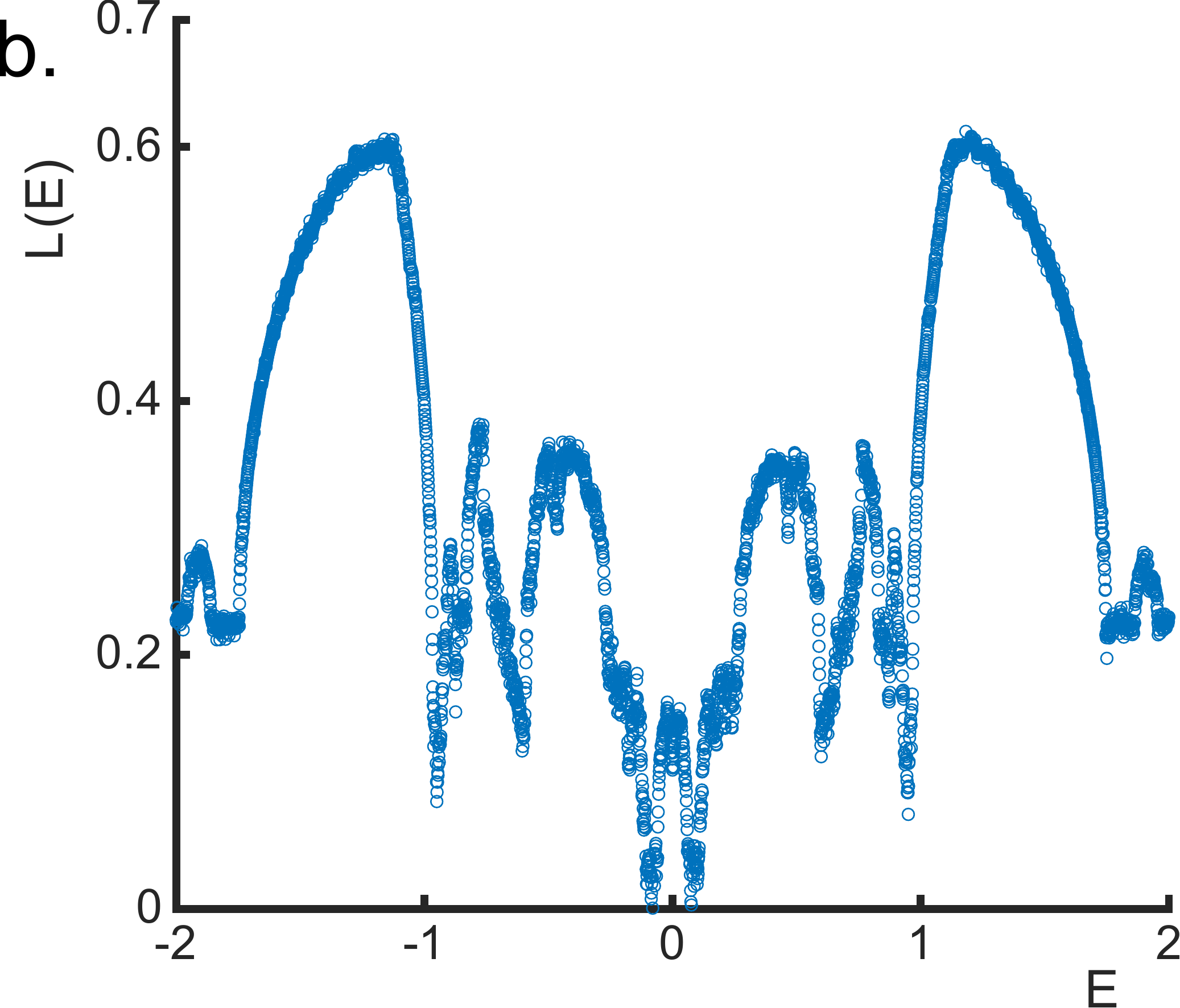}
\caption{(Color online) The Lyapunov exponent as a function
of energy is plotted for canonical $\beta=\frac{\sqrt{5}-1}{2}$
in the insulating regime, $t=0.4V$. Zeeman splitting is chosen to be $\Delta=V$.
In the upper panel, (a), spin-orbit coupling
is absent. In the lower panel, (b), the parameter $t_1$ is chosen to be
$t_1=0.05V$. It is seen that two plots differ only near $E=0$, where small
$t_1$ causes the metallization. Blue and red curves in (b) correspond to two
eigenvalues of the transmission matrix. Energy is measured in the units of $V/2$.}
\label{ft1smallperiod}
\end{figure}

To illustrate our findings, we presented the numerical results for
large modulation periods, $\beta \ll 1$. For these periods the
semiclassical description applies, which allowed us to interpret
the findings in the language of the phase-space trajectories.
%
%Our main result is the delocalization of the Zeeman-split states in the AA model
%due to a weak spin-orbit coupling. This effect can be interpreted analytically
%for long periods of modulation, $\beta \ll 1$, when the description based on the
%phase-space trajectories applies.
In most studies, however, the inverse
%``most irrational"
``golden mean"
value, $\beta=\frac{\sqrt{5}-1}{2}$, is employed. For this period, the localization
properties of the eigenstates in the insulating regime are most complex, in the sense,
that $L(E)$ takes very different values for close energies.
This ``band-structure-induced"
%oscillations
wiggles in $L(E)$ are most pronounced in the vicinity of the transition,
as is illustrated in Fig. \ref{fnumerical}
(black curve). We emphasize that the effect of delocalization due to small $t_1$ persists
for canonical $\beta=\frac{\sqrt{5}-1}{2}$.
In Fig. \ref{ft1smallperiod} we show the curves $L(E)$ for this value of $\beta$ calculated without spin-orbit coupling, $t_1=0$,  and with weak spin-orbit coupling, $t_1=0.05V$,
%spin-orbit coupling.
We see that finite $t_1$ makes almost no difference except for the domain near $E=0$, where it turns the insulator with $L(0)\approx 0.2$ into a metal.

It is instructive to put our main finding into a more general perspective. The
closest analogy to the effect we report can be found in Ref. \onlinecite{PRL1995}.
In this paper the orbital motion of a 2D electron in a strong perpendicular magnetic field was considered.
It was demonstrated that spin-orbit coupling between the Zeeman-split Landau levels
assists the passage of electron through the saddle points of a smooth random potential,
and, thus, facilitates delocalization.

Delocalizing effect of spin-orbit coupling in the quantum Hall regime
is expected\cite{Khmelnitskii} to manifest itself via the splitting
of the extended states in two overlapping spin subbands.
This  is in accord with later
numerical simulations\cite{LeeChalker,HannaQuantumHall+SO}

Concerning the standard physical mechanism  of spin-orbit facilitating of delocalization\cite{Hikami},
it is based on the suppression of  constructive interference of two scattering paths related by time reversal.
In two dimensions it leads to the crossover from weak localization to weak antilocalization in the magnetoresistance
curves. It is inefficient in the problem we studied due to the presence of strong Zeeman splitting.
Note finally, that quantization of the phase-space trajectories in a weak magnetic field
in metals with  strong spin-orbit coupling\cite{Beenakker,Glazman} had recently became a hot
topic in relation to Weyl semimetals.

  %is inefficient in the 1D AA model. The origin of this mechanism is suppression of the constructive
%  interference of two scattering paths related by time reversal

\vspace{4mm}

\centerline{\bf Acknowledgements}
The work was supported by the Department of
Energy, Office of Basic Energy Sciences, Grant No. DE-
FG02-06ER46313.

\vspace{3mm}

\section{Appendix I}
We adopt and extend the numerical procedure, described in Ref. \onlinecite{NUMERICAL}, to calculate the Lyapunov exponent, $L(E)$, for the Hamiltonian Eq. (\ref{eqh1}). Rewriting the Hamiltonian  Eq. (\ref{eqh1}) in the  form similar to that in Ref.  \onlinecite{NUMERICAL} one has
\begin{multline}
\label{AIHamiltonian}
\hat{H}=\sum\limits_n \left(V_nI+\Delta\sigma_z\right) |\phi_n\rangle\langle\phi_n|\\
+T\sum\limits_n \Big( |\phi_n\rangle\langle\phi_{n+1}|+|\phi_{n+1}\rangle\langle\phi_{n}|\Big),
\end{multline}
where $|\phi_n\rangle$ is a $2\times1$ vector
\begin{eqnarray}
\label{A1phi}
\phi_n=
\begin{pmatrix}
\psi_n^{\uparrow} \\ \\ \psi_n^{\downarrow}
\end{pmatrix},
\end{eqnarray}
 and $\psi_n^{\uparrow(\downarrow)}$ corresponds to up(down) spin projections. The matrices $I$ and $\sigma_{z}$ are the identity matrix and the Pauli matrix, respectively.
 The $2 \times 2$ matrix, $T$, has the meaning of the transmission matrix and has the form
\begin{equation}
\label{AIT}
T=\begin{pmatrix}
t && it_1 \\ \\ -it_1 && t
\end{pmatrix}.
\end{equation}
The diagonal terms, $V_n=V\cos\left(2\pi\beta n\right)$, stand for on-site energies.
%is the onsite energy of site given by $V\cos\left(2\pi\beta n\right)$.
Parameters $\Delta$, $t$, $t_1$ are defined in the main text.

The numerical procedure in Ref. \onlinecite{NUMERICAL} is based on
step-by-step decimation of sites achieved by renormalization of energies and
coupling matrix elements for remaining sites.
Since $V_n$ is an even function of $n$, we can restrict consideration to $n\geq0$.

As a first step,  consider three sites $n=0$, $n=1$, and $n=2$.
Elimination of the site $n=1$, results in the following renormalization of the bare on-site energies, $\left(V_n I-\Delta\sigma_z\right)$,  of the sites $n=0$, and $n=2$, as well as  renormalization of coupling, $T_{\s 0,2}$,
\begin{eqnarray}
\label{A20site}
\varepsilon_{\s 0}^{\s 1}(E)=\left(V_0 I-\Delta\sigma_z\right) +T^{\dagger}\left(E-V_1I+\Delta\sigma_z\right)^{-1}T,\nonumber\\
\varepsilon_{\s 2}^{\s 1}(E)=\left(V_2 I-\Delta\sigma_z\right) +T^{\dagger}\left(E-V_1I+\Delta\sigma_z\right)^{-1}T,\nonumber\\
T_{\s 2,0}(E)=T^{\dagger}\left(E-V_1 I+\Delta\sigma_z\right)^{-1}T.
\end{eqnarray}
Renormalized energies, $\varepsilon_0^{1}(E)$, and $\varepsilon_2^{1}(E)$ serve as bare energies in
the subsequent elimination steps. At the second step, the site $n=2$ is eliminated using the rules prescribed by Eq. (\ref{A20site}). Repeating this procedure $N-1$ times, one
arrives to the system of two sites, $n=0$ and $n=N$, with effective on-site energies and effective coupling in the form
\begin{equation}
\varepsilon_{\s 0}^{\s N-1}(E)=\varepsilon_{\s 0}^{\s N-2}(E)+T_{\s 0,N-1}\Big[E-\varepsilon_{\s N-1}^{\s N-2}(E)\Big]^{-1}T_{\s N-1,0}(E).
\end{equation}
\begin{equation}
\varepsilon_{\s N}^{\s N-1}(E)=\varepsilon_{\s N}^{\s N-2}(E)+T_{\s 0,N-1}(E)\Big[E-\varepsilon_{\s N-1}^{\s N-2}(E)\Big]^{-1}T_{\s N-1,0}(E).
\end{equation}
\begin{equation}
T_{\s N,0}(E)=T_{\s 0,N}^{\dagger}(E)=T_{\s 0,N-1}\Big[E-\varepsilon_{\s N-1}^{\s N-2}(E)\Big]^{-1}T.
\end{equation}
In Ref. \onlinecite{NUMERICAL} the Lyapunov exponent is defined as
\begin{equation}
\label{AILyapunov}
L(E)=-\lim\limits_{\s N\rightarrow \infty}\frac{1}{N}\ln |\lambda_{\s N}(E)|,
\end{equation}
where $\lambda_{\s N}(E)$ is the eigenvalue of the effective coupling matrix, $T_{\s 0,N}$. In the presence of the Zeeman splitting and the spin-orbit coupling the eigenvalues are non-degenerate, which results in two Lyapunov exponents. Only the smallest of these two values should be identified with the inverse localization length.

\section{Appendix II}
In the presence of both direct hopping, $t$, and spin-orbit coupling, $t_1$, one can write the tight binding equations for AA model as
\begin{multline}
\label{AIItightbinding}
V\cos(2\pi \beta n)f_n + t (f_{n+1}+f_{n-1})+t_1 (g_{n+1}-g_{n-1})\\
\hspace{55mm}=E f_n, \\
V\cos(2\pi\beta n)g_n +t (g_{n+1}+g_{n-1})-t_1 (f_{n+1}-f_{n-1})\\
=E g_n,
\end{multline}
where $f_n, g_n$ are the amplitudes at site $n$,  corresponding to the up spin and to the down spin. Fourier transformations of $f_n$ and $g_n$ can be written as follows:
\begin{eqnarray}
\label{AIIfouriertransformation}
f_n&=&\sum\limits_m A_m \exp[2im\pi\beta  n]\exp(ikn),\nonumber\\
g_n&=&\sum\limits_m B_m \exp[2im\pi\beta  n]\exp(ikn)
\end{eqnarray}
Substituting Eq. (\ref{AIIfouriertransformation}) in Eq. (\ref{AIItightbinding}) and then comparing the coefficients of $\exp[2im\pi\beta n]\exp(ikn)$ we arrive at
\begin{multline}
\label{AIImodifiedequation}
\frac{V}{2}\Big[A_{m+1}+A_{m-1}\Big]+2t\cos(2\pi\beta m+k)A_m\\
\hspace{17mm}+2it_1\sin(2\pi\beta m +k)B_m=EA_m,
\\
\hspace{-14mm}\frac{V}{2}\Big[B_{m+1}+B_{m-1}\Big]+2t\cos(2\pi\beta m +k)B_m\\
-2it_1\sin(2\pi\beta m+k)A_m=EB_m.
\end{multline}
Multiplying the first equation by $i$ and then adding/subtracting it to/from the second equation yields
\begin{multline}
\label{AIIfinalequation}
\frac{V}{2}\Big[A_{m+1}\pm iB_{m+1}+A_{m-1}\pm iB_{m-1}\Big]\\
+2\
\sqrt{t^2+t_1^2}\cos(2\pi\beta m+k\mp k_0)\Big[A_m\pm iB_m\Big]=E\Big[A_m\pm iB_m\Big],
\end{multline}
where
\begin{equation}
\label{k_0}
k_0=\arctan\frac{t_1}{t},
\end{equation}
and $A\pm iB$ are the new amplitudes. We have reduced the AA model with spin-orbit coupling to two decoupled  AA models for a spinless
electron with hopping amplitude $\left(t^2+t_1^2\right)^{1/2}$. It is important that,
while the eigenvalues of Eqs. (\ref{AIIfinalequation}) are the same for $+$ and $-$
signs, the corresponding eigenvectors are not orthogonal to each other.

\end{document}